# Social Networks and Spin Glasses

*Scott Kirkpatrick\*, Alex Kulakovsky\*, Manuel Cebrian^$, and Alex Pentland^*

*\* School of Engineering and Computer Science, Hebrew University of Jerusalem*

*^ Media Laboratory, Massachusetts Institute of Technology*

*$ Department of Computer Science and Engineering, University of California, San Diego*

## Abstract:

The networks formed from the links between telephones observed in a month's call detail records (CDRs) in the UK are analyzed, looking for the characteristics thought to identify a communications network or a social network.  Some novel methods are employed. We find similarities to both types of network. We conclude that, just as analogies to spin glasses have proved fruitful for optimization of large scale practical problems, there will be opportunities to exploit a statistical mechanics of the formation and  dynamics of social networks in today's electronically connected world.

## Introduction:

Spin glasses [1,2], or rather the techniques of simulation and analysis that have come to be applied to the study of spin glasses, have been a rich source of ideas for managing large and complex engineering optimization problems [3,4,5].  The chief reason for this transfer of knowledge is that the frustration or conflicting interactions in spin glasses which lead to highly degenerate low-energy states and hysteretic dynamics are quite similar to the opposing goals which must be managed in real optimization problems.  Magnetic hysteresis is like the time consuming search for good solutions to such practical challenges.  In recent years, the communications networks and social networks that are created by global scale Web applications have reached scales as large as any logistics or CAD problem.   "Mining" the correlations and predictions possible with such large networks is an advanced and highly profitable art, yet it uses only probabilistic methods, hardly the full scope of statistical mechanics [for a few efforts in this direction, see 6,7,8].

In this paper we explore the differences between two classes of large networks that accomplish the same thing – transporting messages across long distances while minimizing the number of connections required – but the objectives for which the networks are created are quite different.  The internet's physical topography (observed by packet tracing and similar methods) is grown spontaneously under rules which ensure that global connections are achieved.  The CDR or "call detail record" network induced by collecting records of all telephone calls made in one area over a period of time, is driven by our desire to exchange news and keep in contact with our closest friends, and perhaps do business with a network of contacts that reaches a bit further. The first, a classic communications network, has long ranged interactions if viewed as a physical system. The second has been considered a classic social network, the result of only local preferences. We will analyze a large and unusually complete CDR data set to see just what differences emerge. If the form of the interactions makes the CDR network very different from a communications network then it is likely that modeling with physical analogies may give useful insight into the formation of strongly correlated activities, their rate of growth and their eventual limitations.

## Communications networks:

Ad hoc dynamical models have been proposed for the growth or evolution of networks with long tails or even power law distributions of their site degree and other characteristics. The best known of these is Barabasi and Albert's preferential attachment model [9] – new sites join a growing network by making connections with probability proportional to the degree of the site that they connect to. Additional links are formed within the network with both ends attaching to existing sites with probability proportional to their degree. This process yields a power law distribution of site degrees, with an exponent that depends on the relative rates at which new sites and new links between existing sites are added. This model was initially proposed as a generative explanation of the observed long tails in the topography of the Internet, the physical links connecting its subnetworks or "autonomous systems" (ASes).

Data characterizing the AS graph of the Internet is available from several sources, and indeed shows an apparent power law distribution of its site degrees. Shalit et al.[10] first used a technique borrowed from graph theory, k-pruning, to characterize the roles that the sites in such a preferential attachment network model will play. The k-core of a graph is the largest subgraph in which every site has at least k immediate neighbors. This is unique and easy to construct by pruning away, recursively, all sites with fewer than k neighbors (and their links) until only the k-core remains. If one carries out the pruning for each successive integer value of k, starting at k=1 and stopping when no sites remain, the sites that are removed at each value of k (they belong to the k-1 core but are not in the k-core) form a set of k-shells. Shalit et al. found that the sizes of the k-shells in their models follow a strict power law distribution. When the same pruning technique was applied to the actual AS graph data by Carmi et al.[11] (see also work along these lines by the Vespigniani [12] and Stanley[13] groups), the distribution of k-shell sizes, shown in Fig. 1, closely resembles the result of Shalit et al., with interesting differences at the outer edges and in the "nucleus " or innermost k-core. Carmi et al. discuss the history and other implications of this approach for the AS graph. Note the distinct, large "nucleus" set of sites in the innermost k-core in the AS-graph. Although these are found automatically by the k-pruning procedure, they prove to all be known large data carriers, spanning countries and in, some cases, continents. A few of the nucleus sites are companies, such as Google, that need such a presence for their business.

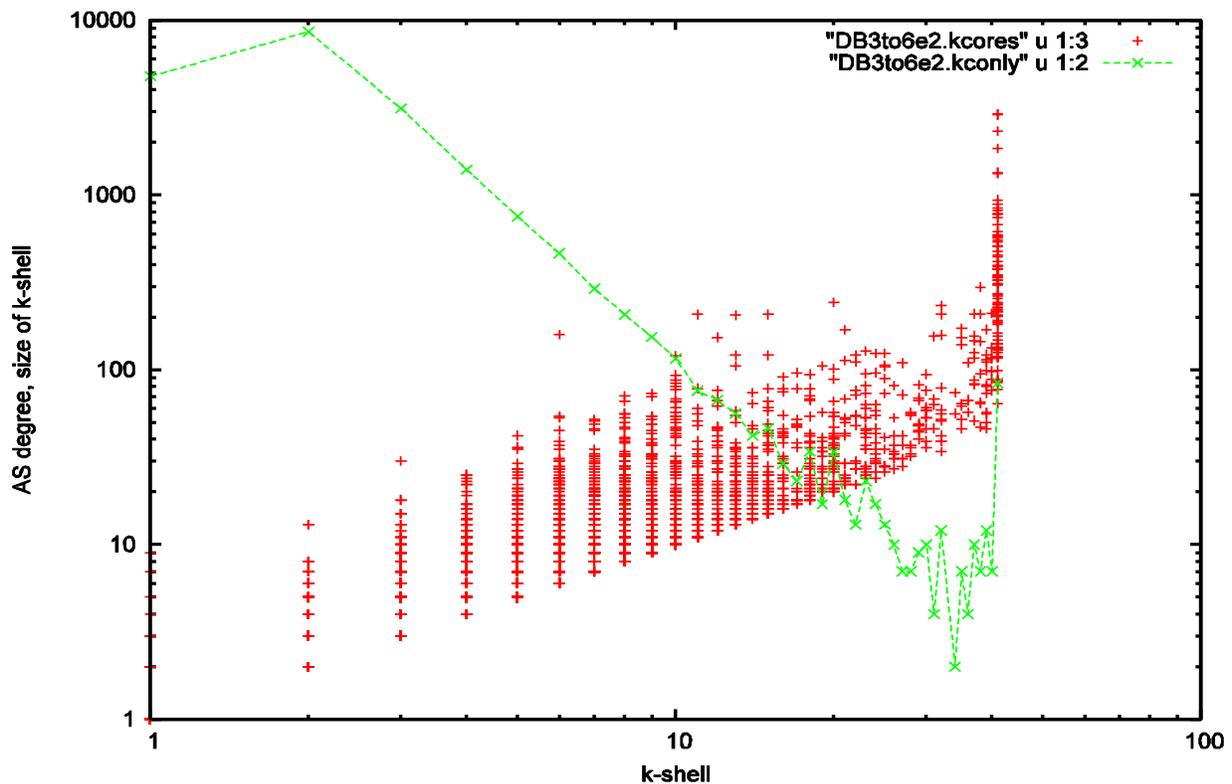

Fig. 1 K-shell analysis of the Internet's AS-graph shows power-law behavior first seen in the preferential attachment models of network formation.

## The CDR data set:

We shall use this same technique on the CDR graph data. We use four weeks' call detail records, including nearly all phone calls placed during August 2005 in the United Kingdom. This data covers more than 90% of the mobile phones and more than 99% of the residential and business landlines in the country. More than 7 billion calls were logged on more than 100 million different numbers, each timestamped with the duration and time of origination accurate to the second. The numbers are hashed to preserve anonymity. Data sets like this, although generally much smaller, have been used in recent publications [14–18]. The calls were then aggregated into links between a specific source and destination, yielding 1.503 billion distinct links.

A clear diurnal variation appears in the data, as shown in Fig. 2. The call volume decreases in stages, from Mondays, to midweek, with a further decrease on Fridays and for each weekend day. This suggests that we can usefully separate the calls into two categories, using their time stamps. We separate calls initiated between 8 am and 6 pm on a work day ("work period") from the others ("leisure time"). Our analysis shows that these two subnetworks of calls and callers have rather different characteristics. August is a popular month for vacations, so this separation may not be absolute. In the full data set there were 1, 213M work period links and 517M leisure period links. 226M links were observed in both time periods, representing 44% of the leisure links. A rather large fraction of the work links (81%) are not seen outside of working hours.

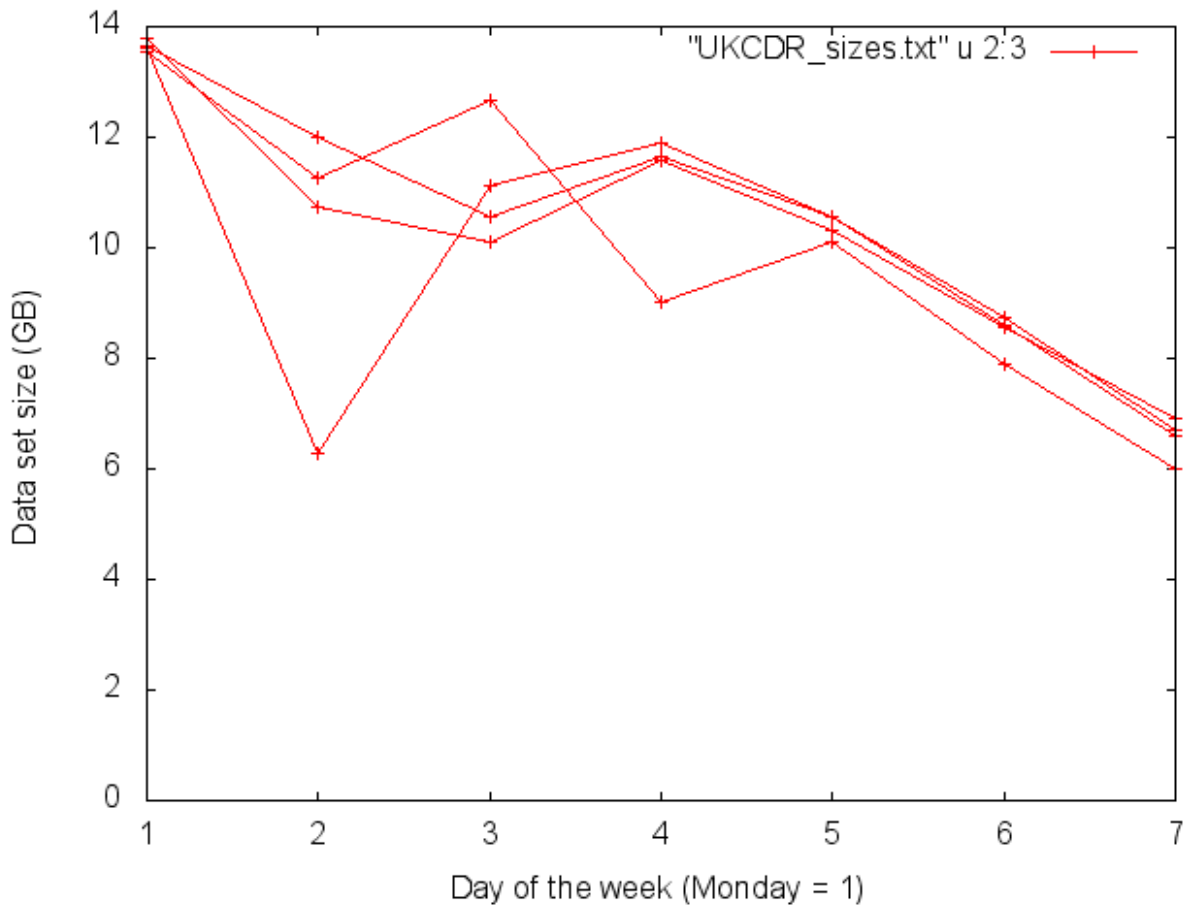

Fig. 2 CDR daily data set sizes for 28 days in August 2005, plotted against day of the week. Diurnal variation in the UKCDR data set size suggests separating work from leisure. The low value of one of the Tuesday data points is due to a partly corrupted file, which is missing data.

Since highly directional links express a different sort of relationship than links in which the calls are reciprocated more equally, it is common in the discussion of social networks to restrict attention to the links in which the smaller of the number of calls from i→j or j←i is at least 1, and to consider these as undirected. The restriction to reciprocal links greatly reduces the size of the network that will be analyzed. We find 166M reciprocated links in our data set, 129M links in the work period, and 72M in the leisure period. The overlap, 35M links, is nearly half of the leisure set, and 27% of the work period links.

A useful and relevant small test set is all the links seen within a single large metropolitan area, coded as "PnLa", since those are the first four letters of the hashed IDs in this area. When we select this area as both source and destination in our hashed addresses, we find 28M links in total, of which 21M are work links and 11M are leisure links. The overlap, 4M links, is again a small portion of the work links, and a larger portion of the leisure links. Fig. 3 summarizes the data size reductions achieved by each of these selections. The line labeled "both periods" lists the number of links that are present in both work and leisure time periods for each data set. A stricter definition of "reciprocated" links, requiring at least four calls in each direction along a link, does not provide much further data volume reduction. We did not pursue this criterion in our analysis.

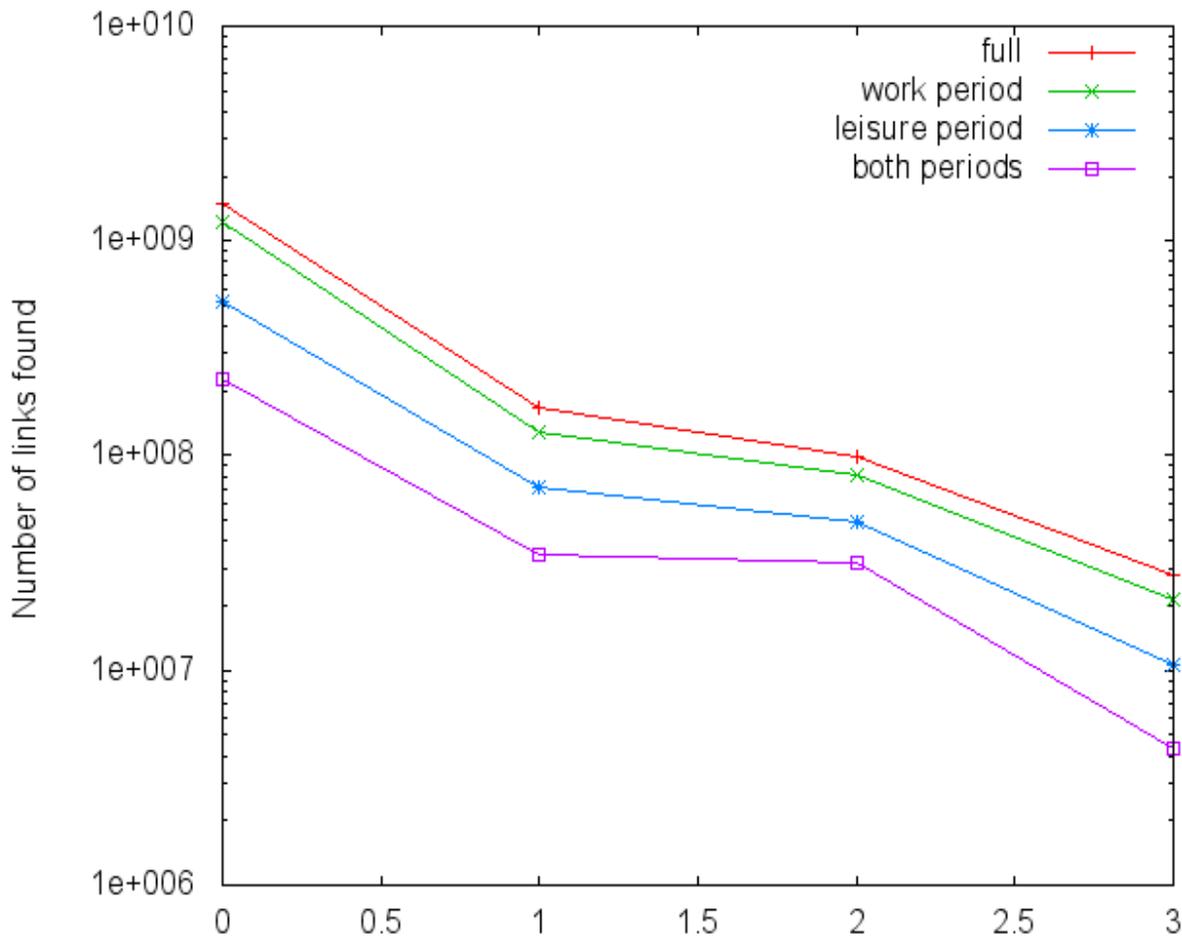

Fig. 3 Telephone call graphs can be reduced in scale, by extracting work or leisure periods, restricting attention to the links in which information flows both ways. We show (left to right) the sizes of the full graph, full graph with only links reciprocated at least once, at least four times, and unrestricted links only within the PnLa metropolitan area.

We used Hadoop in the CDH3 release [19] for a rapid parallel means of aggregating the call logs down to distinct links. We linked it with a distributed memory cache [20] and code that detected and corrected rare race conditions in order to assign a unique numeric index for each telephone number that appears anonymized in the data set. Note that not all IDs are phone numbers in the UK, as foreign destinations and various aggregated addresses will also appear in the call logs.

We assigned 117M distinct IDs in the full data set, 107M of which are seen in the work timeframe, and 88M occur in the leisure data. The overlap in terms of telephone numbers is larger than in terms of links – 78M numbers, nearly 89% of the leisure numbers, participate in both networks. If we restrict our attention to links reciprocated one or more times, 53M of the IDs seen in the full data set participate, 48M of them during the work time period, and 41M during the leisure time. Again, the overlap, 37M IDs, is quite large. Finally, 3M distinct IDs are present in the PnLa-only data set. There were 2.8M ids in the work data set, 2.4M ids in the leisure data set, so 2.M ids are present in both timeframes.

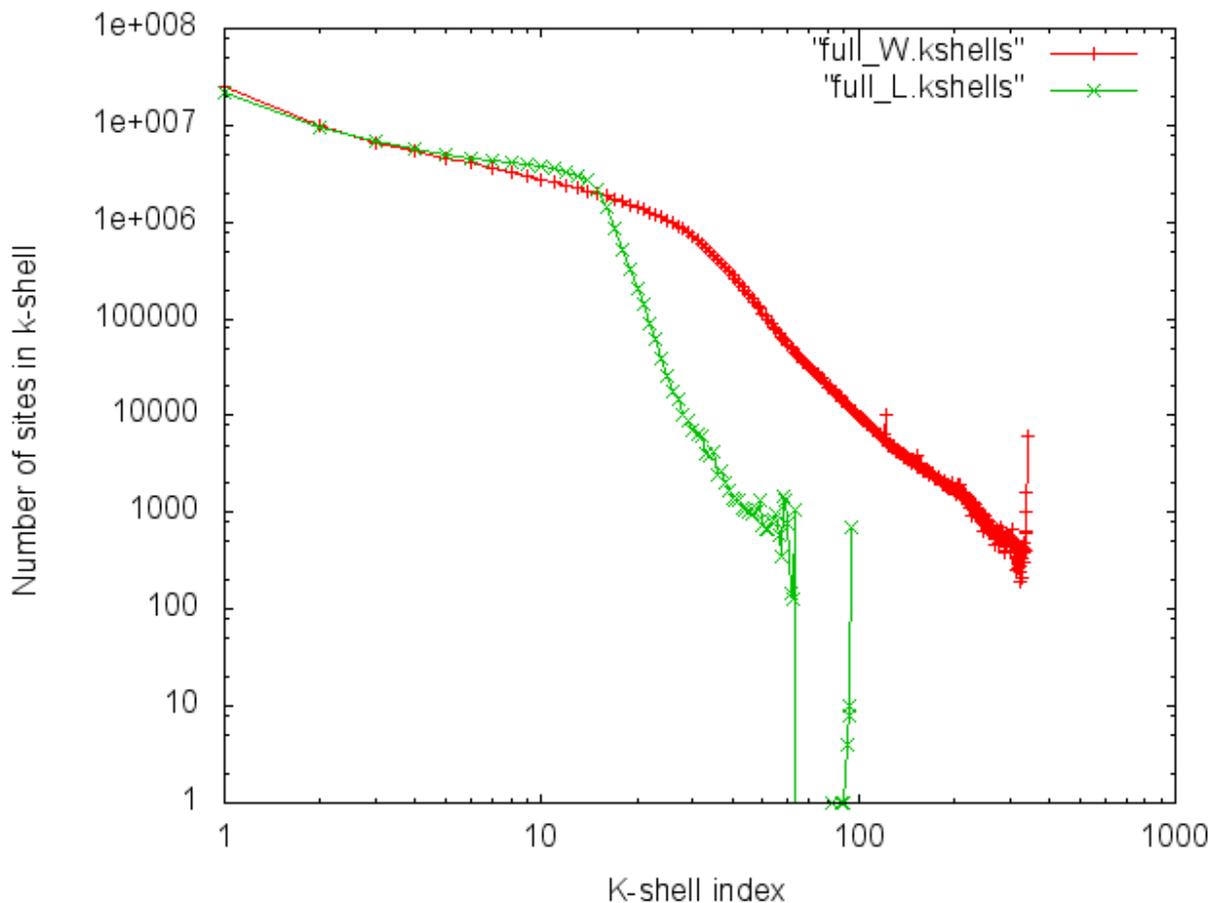

Fig. 4  K-shell size distribution for the full UKCDR data set, unrestricted links, separated into those seen during work hours and those seen at all other times during the month.

## K-pruning analysis of the CDR data set:

The K-shell size distributions for the work and leisure data sets in the full UK network are plotted in Fig. 4. The shapes of these k-shell size curves are very different from the straight line behavior seen on log-log plots with communication networks such as the AS graph of Fig. 1 and that studied in [10].  The note by Cebrian, Pentland and Kirkpatrick [21] discusses these differences and interprets them as due to the different role of three different levels of a communications hierarchy.  The initial flat characteristic might  be the result of local connections between neighbors and close friends, as this is most of the leisure time data set.  The sloping region with a more power-law like decrease, also seen in the AS graph, then represents calling patterns that connect local regions, and may be the result of the communications from and between companies, and other work-related exchanges of information.  There is a possible role for organizations on the national scale, which may be seen at the highest values of k in the nucleus (or of site degree).  Whether these are government or commercial enterprises is unknown.  We also note, for further discussion, that the leisure k-pruning actually exposes two nucleus subgraphs, while the work k-shells show a single nucleus as an endpoint but one or more spikes along the deeper parts of the distribution, that may reflect different functions than the shells before or after.

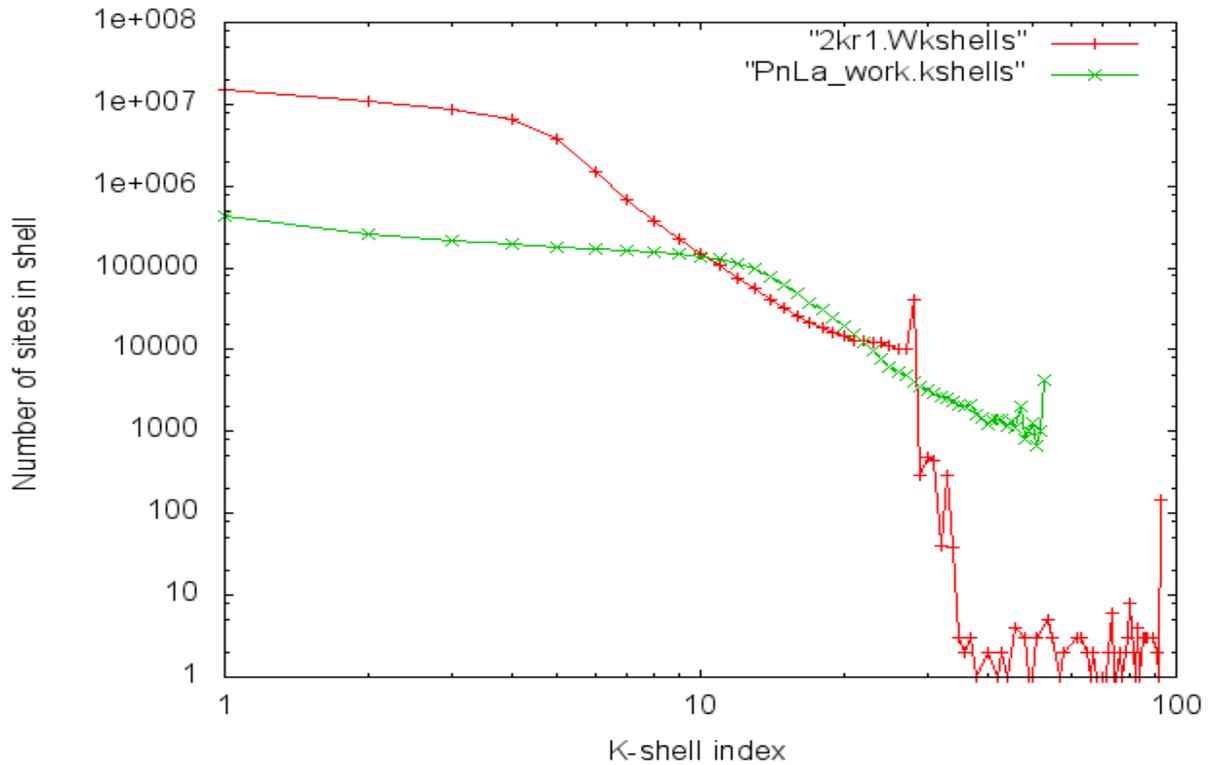

Fig. 5 K-shell distributions during the work time period, singly reciprocated links from the full network, and all links from the PnLa metro region.

Figs 5 and 6 show the k-shell distributions for the full network, when only reciprocated links (at least one call in each direction) are included and for the network within the PnLa metropolitan region, using all links seen. The two curves in Fig. 5 resemble the curve of the work period data in Fig 4, while the leisure data for the reduced data sets in Fig. 6 exhibit the flatter initial region and steeper decline at large k that is seen in the leisure data of Fig. 4. The reciprocated data for the work period shows a small deep core that splits off from the rest of the distribution as if it consists of something unrelated to the rest of the network, a question that we will explore below. The leisure data in each case has a lower density of links, and reaches much smaller values of k, than the work data. The flatter initial curve suggests that this means that the leisure network is more concentrated in the small k regime (fewer contacts) rather than just that it is a result of having fewer links. The split-off inner core seen in the full network's leisure network is not so evident in the two reduced networks for the leisure period, although there are hints of several maxima seen in the shell size plots near the nucleus.

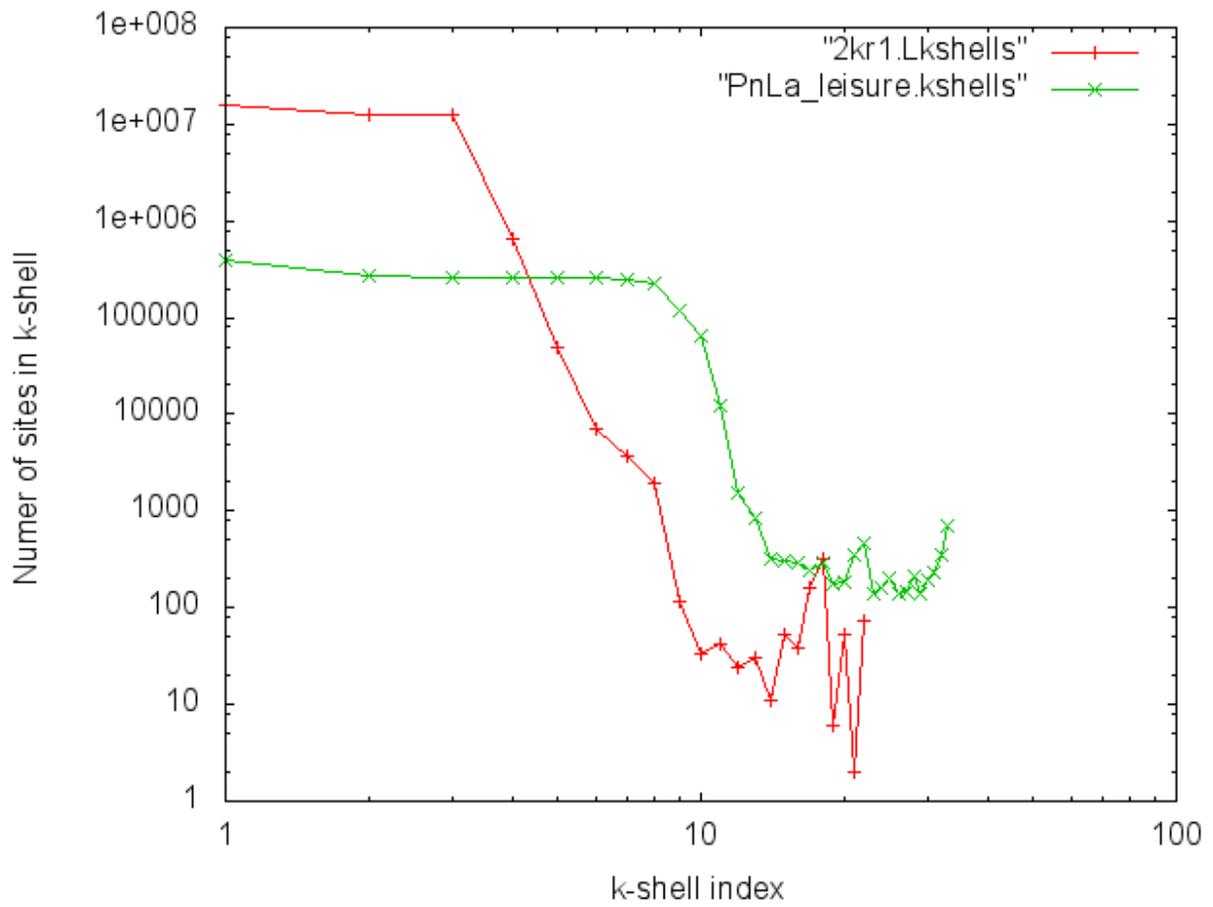

Fig. 6  K-shell size distribution during the leisure periods, singly reciprocated links in the full network, and all links in the PnLa metro region.

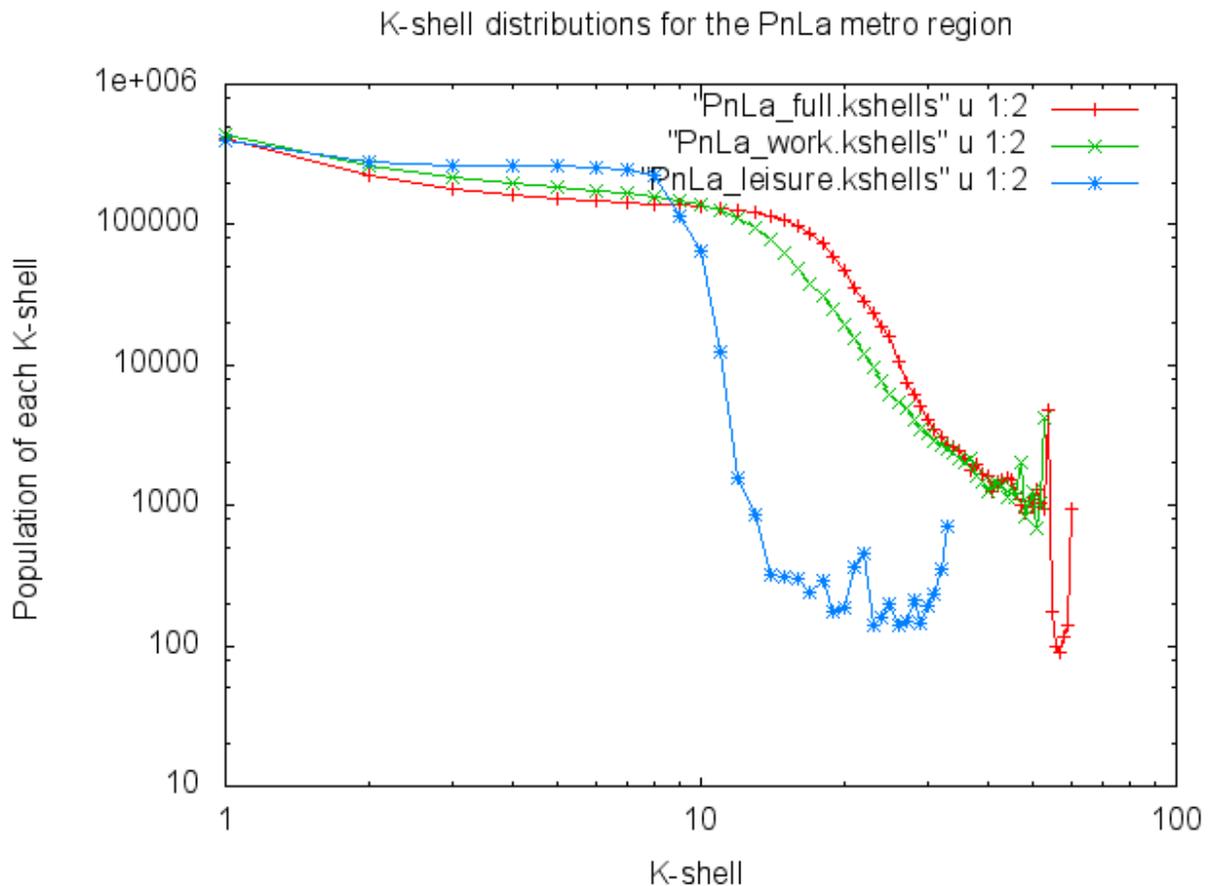

Fig. 7 K-shell sizes in the full PnLa data set, its work and its leisure components. The loglog scale emphasizes the high k tail of the distributions.

Figs. 7 and 8 contrast the two time periods with the k-pruning results for the full PnLa data set, and show the effect that the style of the plot has on the impression we extract. In Fig. 7 we see the much steeper falloff in the tail in the leisure time period, and the fact that the work data almost completely accounts for the high connectivity (high k) shells of the full network. In each of the three curves it appears that the deepest k-cores come from multiple causes, giving rises to several peaks. Fig. 8 uses a linear scale to highlight the small k behavior of this network analysis. The leisure data in Fig. 8 shows that the sites which in the work period exhibit higher values of k are squashed down into a series of shells with k<10 and populations roughly twice that of the early parts of the work period and full distributions. The work period has a gradual and probably "fat" tail to the high k regime, while the leisure period seems almost cut off entirely at values of k much larger than 10. Note that both work and leisure periods have almost identical populations of sites in shells k=1 and 2, with differences beginning to appear at larger values of k.

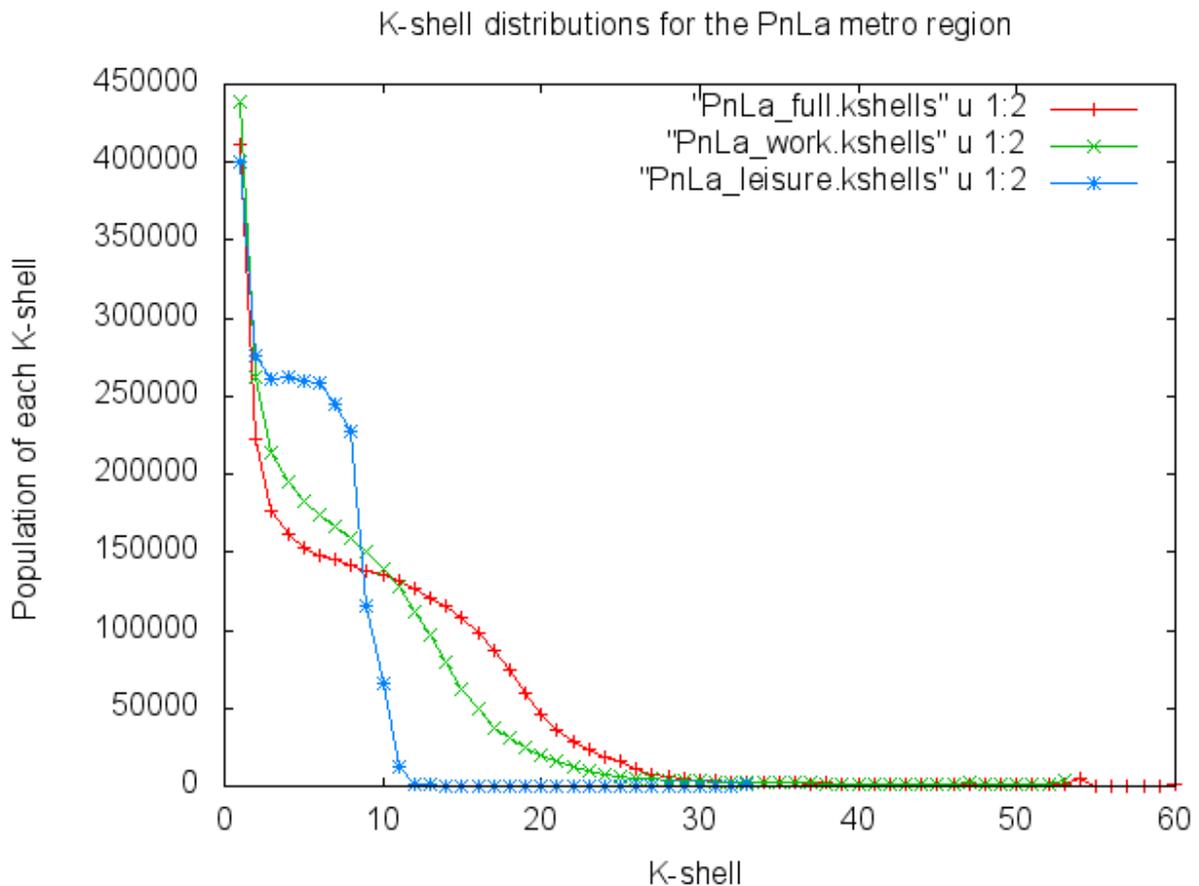

Fig. 8  K-shell size distributions for the PnLa data, linear scale to emphasize the differences at small k.

In order to get a better sense of what roles the links in the CDR network are playing, which represent people talking to people, and which reflect business communications, commercial interests, or national scale services, we will characterize the links in the network grouped by the k-shells which they connect.  Fig. 9 shows the number of links connecting to each shell in the full work and leisure datasets.  The largest numbers of links occur, not at the edge of each network (k=1), but at modest values of k.  Because the decrease in the number of sites in each shell is slower than 1/k (after k = 1 or 2), this gives a  smoothly rising curve for each data set's link population per shell in Fig. 9.  For the full leisure network, we selected shells k=1, 7, 14 and 28 for further study.  For the full work network, we select k=1, 10, 20, and 40.  There also appear to be special clusters with high connectivity embedded in certain k-shells.  In the full work data set these create spikes at k=122 and k=267.  Finally, the deepest k-core, or the nucleus of the full work data set lies at k=341.

In the full leisure data set, the deepest nucleus lies at k=95, but does not seem to involve an unusually large number of links, while the earlier endpoint at k=63 has several million links to the rest of the network.  We examined the set of sites in the shells k=90-95 which are separated from the rest of the distribution and found that all of them had of order 100 neighbors at the outset of the k-pruning. Most of these links must therefore lie within the cluster and are not made to much smaller k k-shells. It appears likely that the true nucleus of the leisure network is the k=63 shell.

In other cases that we examined we find a significant difference between this network data and the communications networks that we have examined – the sites of highest degree do not appear deep in the network, but are pruned much earlier. In the reciprocated network seen during work hours, the highest degree site had 23,960 neighbors (equivalent to calling 800 numbers a day, likely an autodialer). It was found in shell k=28 out of 92. In the reciprocated leisure network, no site in the shells k>10 had a degree exceeding 10,000, but a few in the earlier shells did.

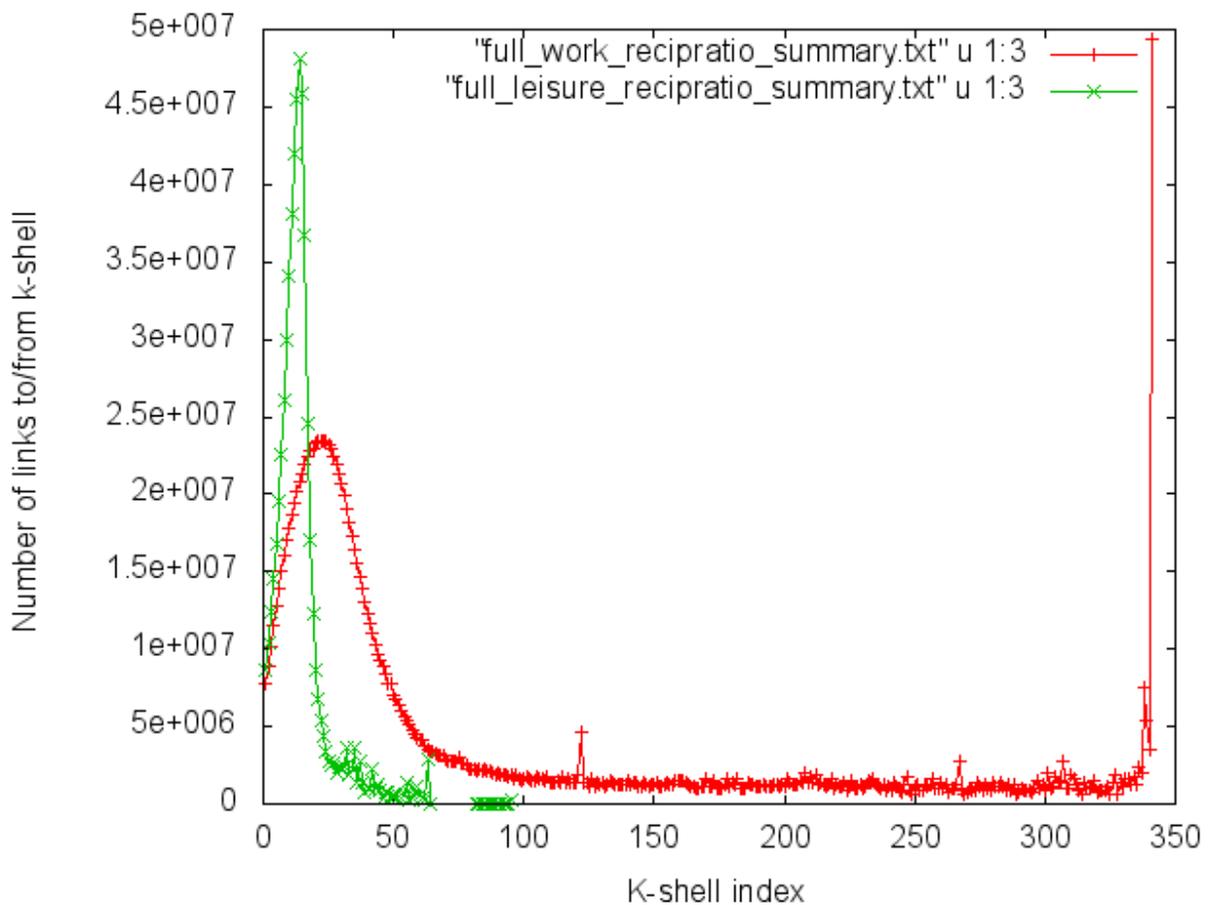

Fig. 9 Number of links connecting to each k-shell in the full data set, separated into work and leisure periods. Note the spikes in the work data set at k = 122, 267, and the final nucleus at k = 341.

The first clue that we can examine to see where in this decomposition the different possible communities in the country under study are served is the distinction between one-way links and links over which the calls are reciprocated. In Figs. 10 and 11 we have plotted the fraction of the links connecting each k-shell to the rest of the network which are reciprocated as a function of the k-shell index. Fig. 10 shows the results for the full work and leisure data sets, while Fig. 11 does the same for the two periods using only the PnLa metropolitan region data set.

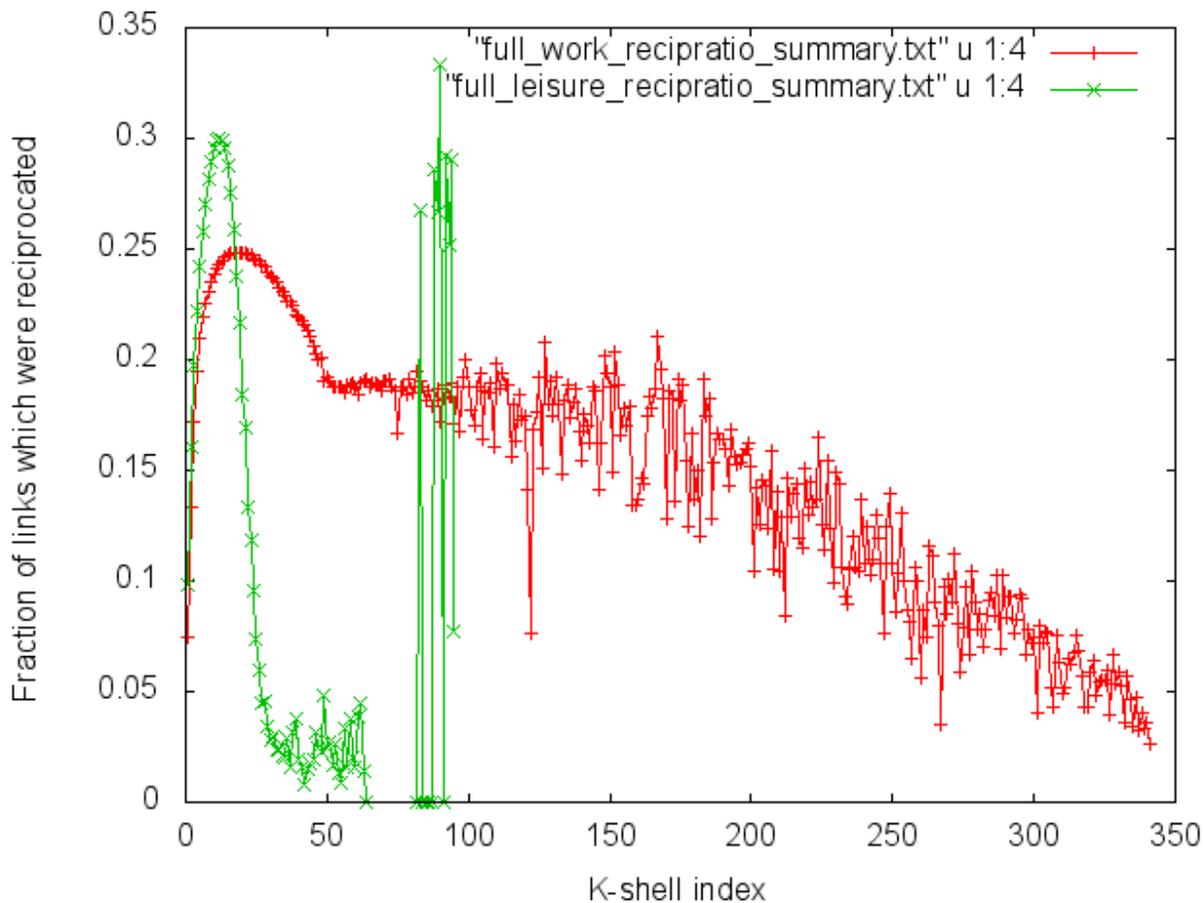

Fig. 10 Fraction of links which are reciprocated, averaged over an individual k-shell, for the work and leisure periods in the full data set.

We see similar behavior in both the full and PnLa data sets. It is surprising that in all four curves the earliest k-shells show a very low fraction of calls which are reciprocated, increasing steadily as the k-shell index increases to about 10 in the full leisure period data set or about 20 in the full work period data set. The same shape at low k is seen in the data reduced to only the PnLa region. These are the k-shells that we have interpreted as due to local interactions, within neighborhoods or cities, among callers who know each other. It comes as a surprise that the very small k-shells, from k = 1 to about 5 or so, are more likely to be unreciprocated, but the decline in the fraction reciprocated beyond the peak at modest values of k might be associated with an increasing fraction of business or institutional calls and numbers, which because of their more hierarchical nature would be more likely to have a one-way character. The fraction reciprocated at values of k > 10 in the full work period network varies considerably more, but continues this steady decline. The variation is largely due to the fact that, as Fig. 9 shows, this is the tail of the distribution, and these shells are quite sparsely populated. In the isolated large k cluster that we see in the full leisure period data set of Fig. 10, the fraction of reciprocated links is again rather high, but this cluster appears to be a relic of some structure present in the work data set, in which the numbers are highly connected to each other but disconnected from the rest of the country.

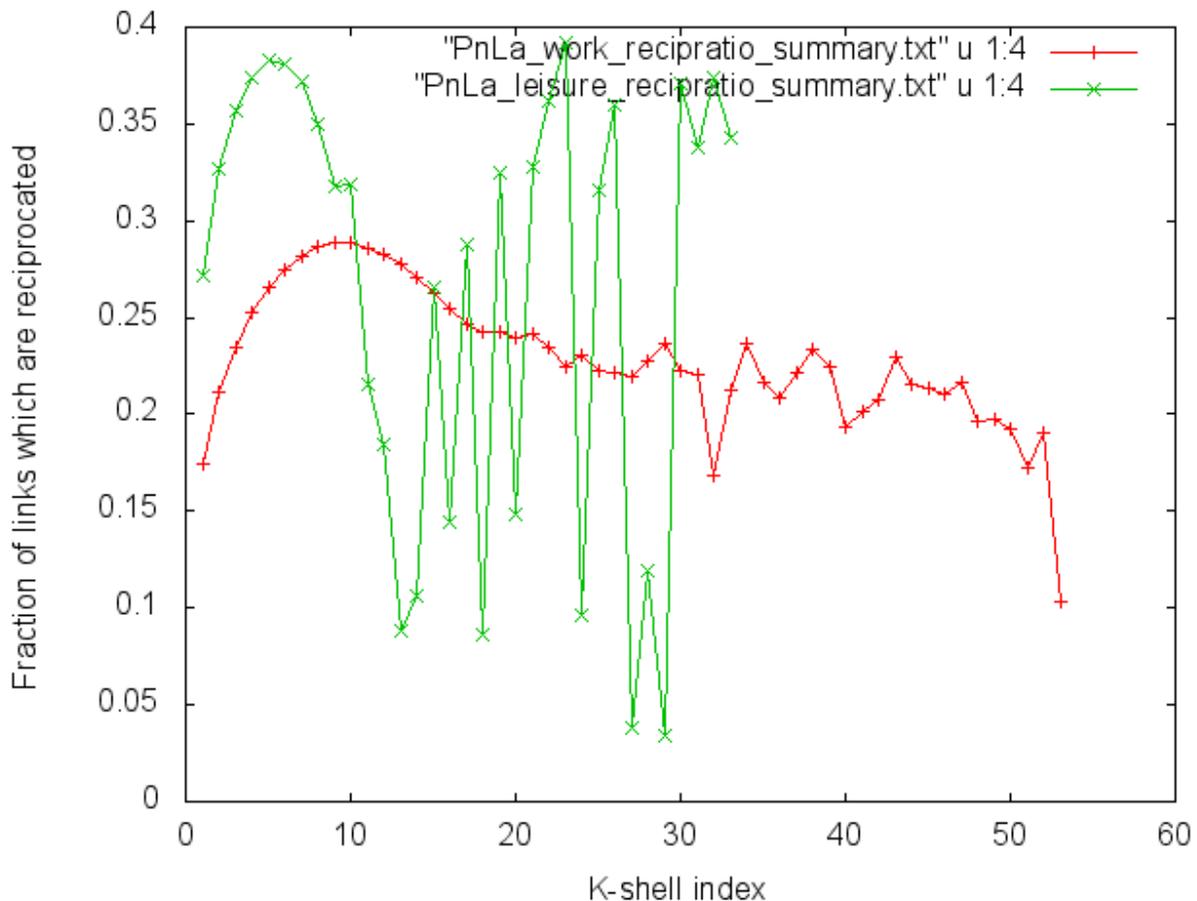

Fig. 11 Fraction of links which were reciprocated, in the PnLa regional data set. Averages over individual k-shells.

The same pattern occurs in Fig. 11, which restricts the data to the PnLa metropolitan region. Each curve has a broad maximum at modest values of k, followed by a slow decrease (in the work period data) at larger k or a sharp decrease (the PnLa leisure dataset) at higher k in the leisure period. Also in the leisure period there is a sparse but reciprocally interconnected set of sites at the higher values of k. Since the PnLa leisure data set is rather small, we won't speculate further on how this high-k cluster comes about in a single city.

Figs. 12 and 13 refine the analysis of the reciprocated ratio by plotting for the leisure period (Fig. 12) and for the work period (Fig. 13) in our full data set, the fraction of links from a given k-shell that are reciprocated, as a function of the index of the other k-shell to which the link connects. The shells which are profiled in these figures are the k-shells which comprise the broad peak in which most of the population of sites (phone numbers) reside, plus the values which showed up as spikes in plot of k-shell site population in Fig. 9. Each of these spikes may indicate the presence of a potential embedded nucleus.

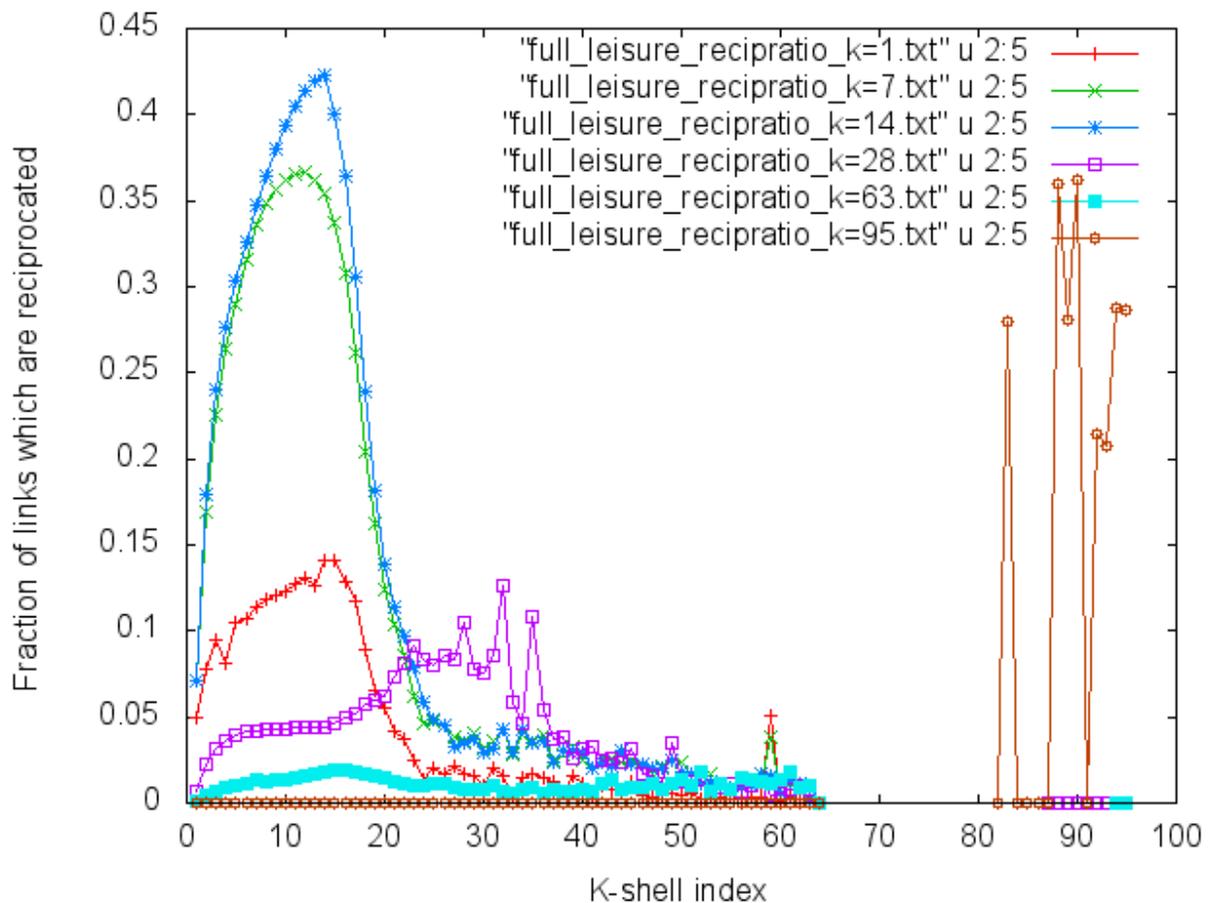

Fig 12. For each of several k-shells in the full leisure CDR network, the fraction of the links to each of the other k-shells which were reciprocated. K=1, 7, 14, and 28 are plotted, along with the two "nuclei," k=63 and 95. Links between the shells k=80-95 are highly reciprocated, but there are no links between that cluster and the lower k (k<65) shells.

The analysis of Fig 12 shows up the differences between the three clusters found deep in the full leisure network. The nucleus of sites in the k=95 shell connect only to the other sites with k > 82, and to no earlier shells. Their fraction of reciprocated connections is quite high. The nucleus shell at k = 63 connects to all the shells, but only a few percent of the links to any other shell are reciprocated. The three shells at k = 7, 14, and 21, which sample the earlier k-shells show similar behavior, a low fraction of links reciprocated at first, and then a high fraction in the middle of the apparently "social" k-shells from 3 or so to 20. The final curve for k = 28 at the upper edge of this portion looks different. It makes a larger fraction of reciprocated links within a region from k = 25 to 35 that the earlier k-shells seem to have ignored.

Fig. 13 shows the same comparison for a relevant set of k-shells in the work data set. The nucleus shell at k = 341 behaves similarly to the nucleus site at k = 63 in the leisure data set, making mostly one-directional links to all the earlier k-shells. The other two possible embedded clusters, at k = 122 and 267 in the work data set, behave differently, making a large fraction of reciprocated links to shells at depths in the middle or deeper in the network's k-shell structure. The k=10, 20 and 40 data also makes reciprocated connections to this deep region, and in addition shows 30-40% reciprocated links among the early k-shells. Finally, we plot the fractions of the links to the k=1 shell that are reciprocated and find a pattern very similar to that of the nucleus. The fact that the very first k-shells behave differently from the shells with moderate values of k is very different from the behavior seen in a general purpose communications network, e.g. the Internet AS graph.

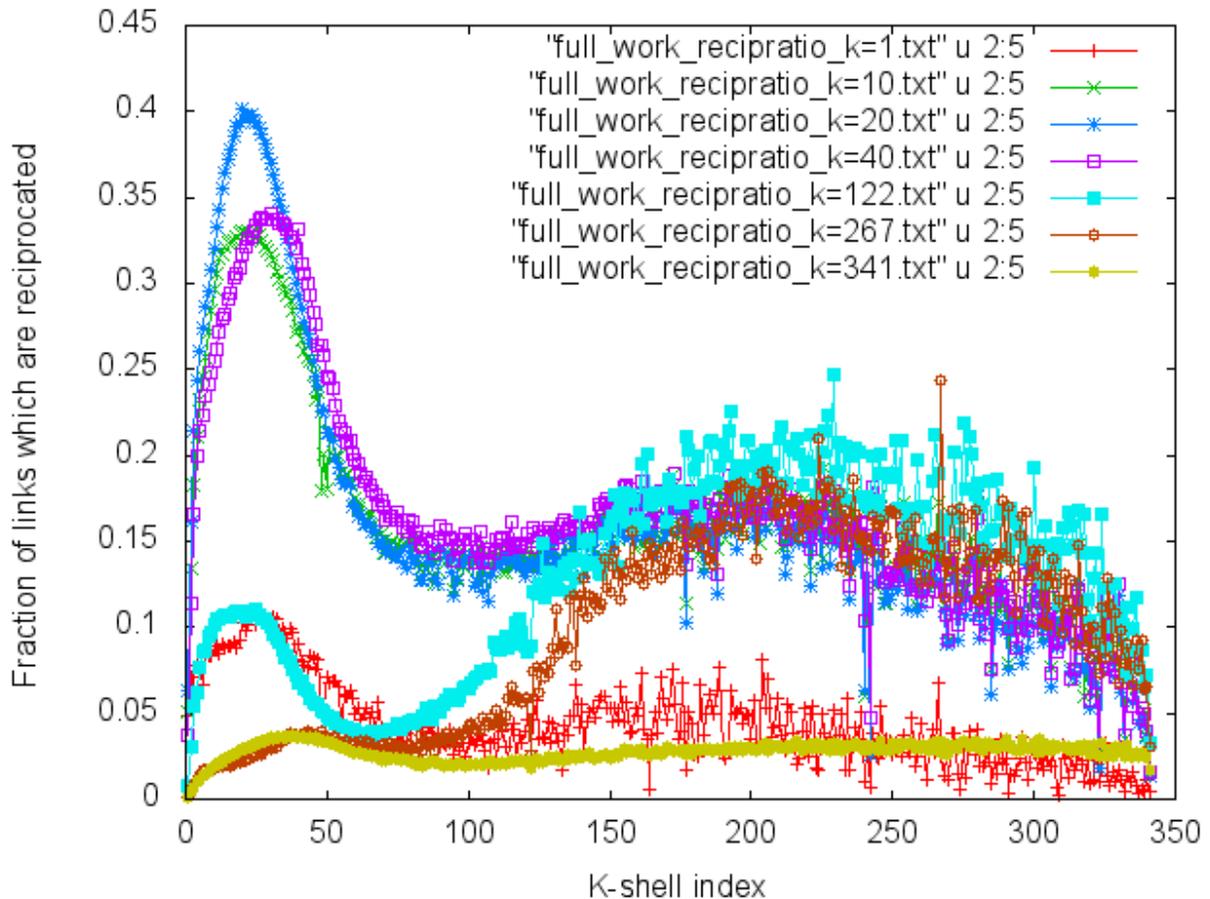

Fig. 13. For several k-shells in the full work period CDR network, we plot the fraction of links to each other k-shell that are reciprocated. K= 1, 10, 20, and 40 are plotted, along with the k-shells with spikes in their population at k= 122, 267, and 341.

## Other analyses:

A classic analytical tool for distinguishing social from communications networks, is the degree-degree correlation. In the networks formed by collaborations between scientists or actors, in which a link is a piece of work that two people have collaborated on, sites with high degree have neighbors with high degree on the average, while sites with few neighbors have neighbors of low degree. In communications networks, on the other hand, the reverse is seem. Sites with low degree tend to connect to sites of much higher degree, and sites with high degree tend to be surrounded by sites of much lower degree. This second pattern, called disassortive in Newman's reviews [22], could be the result of incentives to spread messages as quickly and broadly as possible. It is evident in the Internet AS graph. Studies of social networks, usually much smaller in extent, show an upwards-sloping plot of degree-degree correlation. Analyzing our CDR networks in this way, we see both behaviors. For small degree (which roughly corresponds to small k), we see an increase in the average degree of a neighbor as the degree of a site increases, yet for the larger degrees, the average neighbor degree increases strongly.

This is seen in the full plots for both leisure and work periods, and in the PnLa data just as in the full data set.  We believe the low degree (and low k) nodes of the CDR network represent people communicating with friends and on a local scale.  The high k and high degree sites involve more contacts within the space of a month  than most people would be expected to know personally, so these links should be the result of institutions, business activities and perhaps government services.  Our conclusion is that the personal telephone network shares the characteristics of traditional social networks [22], but the more hierarchically organized institutional links  in the CDR network have a geometry like that of other  communications networks.

One result of the degree-degree correlation analysis proved surprising.  The curves were flat when plotted for the network formed from the full data set restricted to links reciprocated at least once.  At present we have no explanation for this ccounter-intuitive behavior.  The simplest expectation, that reciprocated lnks represent people-to-people communications, would have predicted an increasing degree-degree correlation curve, as is seen in smaller social networks.

## Conclusions:

Telephone call detail record networks have a rich structure, and cannot be simply characterized as social or communications networks, since the networks of several communities that use the telephone for very different purposes are merged together in this data.  In the large scale structure we see correlations that meet community needs on the scale of personal interactions as well as hierarchically ordered linkages that distribute information over long distances, but reach fewer individuals.  Comparing leisure time connections with the larger set of connections made during working hours separates out some of the threads.   Since the network appears to be a superposition of several communities, interacting on their appropriate scales, constructing a predictive model of the growth and dynamical response of the telephone network to disturbances appears to require multiple submodels, each with appropriate but different "physics."

## Acknowledgements:

This work was initiated during SK's summer visit to the MIT Media Lab and its Human Dynamics group during 2010, and continued in summer 2011.  At the Hebrew University, SK and AK have enjoyed the support of the LAWA project, an EC collaborative research project (number 258105) on "Longitudinal Analytics of Web Archive Data."  MC is supported by the National Science Foundation under Grant No. 0905645.   AP is partially supported the ARL under Agreement W911NF-09-2-0053, and by AFOSR under Award Number FA9550-10-1-0122.

**References:**

1. S. F. Edwards and P. W. Anderson, J. Phys. F Met. Phys. **5**, 965-974 (1975).
2. D. Sherrington and S. Kirkpatrick, Phys. Rev. Lett. **35,** 1792-96 (1975).
3. S. Kirkpatrick, C. D. Gelatt, Jr., and M. P. Vecchi, Science **220**, 671-680 (1983).
4. S. Kirkpatrick and B. Selman, Science **264**, 1297-99 (1994).
5. R. Monasson, R. Zecchina, S. Kirkpatrick, B. Selman and L. Troyansky, Nature **400**, 133-137 (1999).
6. R. Albert and A.-L. Barabasi, Revs. Mod. Phys. **74**, 47-97 (2002).
7. J. Park and M. E. J. Newman, Phys. Rev. E **70**, 066117 (2004).
8. R. Pastor-Sarras, J. M. Rubi an A. J. Diaz-Guillera, eds., *Conference on Statistical Mechanics of Complex Networks,* Sitges, Spain, (Springer 2003).
9. A.-L. Barabasi and R. Albert, Science **286**, 509-12 (1999).
10. A. Shalit, S. Kirkpatrick and S. Solomon, poster at "Aspects of Complexity and its Applications, Rome, Sept 2002 (available on www.cs.huji.ac.il/~kirk).
11. S. Carmi, S. Havlin, S. Kirkpatrick, Y. Shavitt, and E. Shir. cond-mat/0601240. Shortened version appeared in PNAS **104**,11150-50 (2007).
12. J.I. Alvarez-Hamelin, L. Dall'Asta, A. Barrat, and A. Vespignani , Networks and Heterogeneous Media, **3**, 271-293 (2008) .
13. M. Kitsak, L. K. Gallos, S. Havlin, H. A. Makse and H. E. Stanley, Nature Physics (2010)
14. W. Aiello, F. Chung, and L. Lu, Proceedings 3$^{2nd}$ Annual ACM STOC, 171-180 (2000).
15. 15 W. Aiello, F. Chung, and L. Lu, in J. Abello, P. M. Pardalos, M. G. C. Resende (eds.), *Handbook of Massive Data Sets*, 97-120 Kluwer, Dordrecht (2002).
16. M. Seshadri, S. Machiraju, A. Sridharan, J. Boulot, C. Faloutsos and J. Leskove, Proceedings of the 14$^{th}$ SIGKDD Conf. 596-604 (2004).
17. N. Eagle, A. Pentland, and D. Lazer, PNAS **106**, 15274 – 15278 (2009).
18. N. Eagle, M. Macy, and R. Claxton, Science **328,** 1029 (2010).
19. T. White, *Hadoop, the Definitive Guide* (Yahoo Press, 2010).
20. R. I. Fitzpatrick, Linux Journal **2004**, 5 (2004).
21. M. Cebrian, S. Kirkpatrick, and A. Pentland, poster at Workshop on Information in Networks (WIN 2010) Sept 2010, ArXiv: 1008.1357.
22. M. E. J. Newman, SIAM Review **45**, 167-256 (2003).